# Broadband on-chip SiN lasers

Ken Liu *
College of Advanced Interdisciplinary Studies, Hunan Provincial Key Laboratory of Novel Nano Optoelectronic Information Materials and Devices & Nanhu Laser Laboratory, NUDT, Changsha, China.
* kenliu@nudt.edu.cn

Broadband active materials are pivotal for advancing emerging technologies spanning on-chip optical interconnects, artificial intelligence, quantum systems and precision metrology. Current semiconductor gain media face bandwidth limitations; and Ttitanium-doped sapphire (Ti:sapphire), the most widely used broadband light-emitting material, covering the red to short-wave near-infrared (SW-NIR) spectrum, lacking emission in the entire visible range. Here, a mechanism for generating ultra-broadband gain is revealed, which utilizes defect and band-tail states in the bandgap, and balances cavity enhanced reabsorption and radiation. By leveraging this mechanism, the gain of on-chip integrated silicon nitride (SiN) is greatly enhanced at longer wavelengths, thereby achieving broadband emission, from blue light to SW-NIR (approximately 450 nm to 1000 nm), and mode-hop-free tuning about 1.6 nm at about 738 nm and amplification at about 532.3 nm was achieved. By leveraging the maturity, cost-effectiveness, and CMOS compatibility of SiN photonics, this work transitions SiN from conventional passive photonic material to ultrawide-band active medium, establishing a disruptive foundation for next generation visible and SW-NIR integrated photonic platforms.

.

## Introduction

The growing demands of integrated photonic systems—spanning optical interconnects, optical computing, artificial intelligence, and quantum technologies—have intensified the pursuit of compact ultra-broadband light sources [1–5]. While supercontinuum generation offers a potential solution [6,7], it relies on materials with nonlinearity, precisely engineered dispersion profiles, and has low power at individual wavelength. Alternative approaches leveraging SiN microresonator-based optical frequency combs [8,9] have unique advantages, but they also face certain challenges: these nonlinear processes demand ultra-high cavity quality factors (Qs), strong third-order nonlinear susceptibility, and precision in dispersion control.

In contrast, on-chip integrated broadband lasers offer a more direct and efficient solution. While passive on-chip SiN resonators have been used to achieve widely tunable lasers [5], their active gain mediums typically rely on semiconductor materials, which are inherently limited by their band structures and gain bandwidth. To address this limitation, SiN has been explored as an active gain medium through rare-earth ion doping [10], a method also applied to other platforms such as lithium niobate waveguides [11]. However, rare-earth ions are constrained to specific emission wavelengths, for instance, erbium ions are utilized for 1.55 μm optical amplification. Among gain materials, Ti:sapphire stands out for its exceptionally broad emission spectrum, ranging from 650 nm to 1,100 nm, making it highly valuable for numerous applications. Recent advancements have demonstrated the successful hybrid integration of sapphire with silicon platforms [12, 13]. Despite this progress, Ti:sapphire's emission spectrum does not extend into the entire visible range, and its integration with silicon poses certain challenges, these include the need for bonding techniques and polishing due to sapphire's extreme hardness, as well as the complexity of etching sapphire using chlorine-based gases, which is not compatible with CMOS processes.

In this work, SiN as a broadband active material was experimentally demonstrated. With a refractive index ranging from 2.0 to 2.2, significantly higher than that of Ti:sapphire (1.76), SiN strip waveguide provides stronger optical confinement. A 12 μm diameter SiN microring achieves emission spanning from 500 nm to 900 nm, while a 36 μm diameter microring extends the spectrum from 450 nm to nearly 1,000 nm, delivering a gain bandwidth exceeding 360 THz, which is about twice that of Ti:sapphire. This work highlights the potential of SiN for on-chip solid-state lasers operating across the visible to SW-NIR spectrum, bridging the emission gap of Ti:sapphire between 450 nm and 650 nm and enabling optical amplification. Additionally, SiN's lasing compared to optical frequency combs eliminates the need for extremely high Qs to achieve broadband emission. Furthermore, the SiN photonic platform is fully compatible with CMOS processes, making it a pivotal technology for advancing next-generation on-chip photonic applications.

## Photoluminescence (PL) of SiN films

Although crystalline silicon nitride possesses a wide bandgap (~4.6 eV) that theoretically renders it transparent in the visible spectrum—incapable of absorbing or emitting light within this range—the optical behavior of SiN films is critically modulated by deposition conditions. In amorphous or polycrystalline SiN grown via plasma-enhanced chemical vapor deposition (PECVD) or low-pressure chemical vapor deposition (LPCVD), deviations from stoichiometry (e.g., variations in nitrogen-to-silicon ratios), the incorporation of silicon nanoparticles or nanocrystals, oxygen and hydrogen doping effects, band-tail recombination processes, structural vacancies, and fabrication-specific parameters collectively modify the electronic band structure, introduce impurity energy level, enabling distinct from the crystalline phase [14–16]. These compositional and structural heterogeneities highlight the tunable optoelectronic properties of SiN thin films despite their nominally insulating nature.

Compared to PECVD-grown SiN, LPCVD-grown SiN is typically denser and more reliable. Therefore, all luminescence studies on SiN in this work utilize LPCVD-grown SiN. Fig. 1 presents the PL spectra of three types of 300-nm-thick SiN films grown under different LPCVD conditions. During LPCVD growth, the precursor gases dichlorosilane ($SiH_2Cl_2$) and ammonia ($NH_3$) are introduced with flow ratios varying from 4:1 to 1:4, and the growth temperature ranges from 750°C to 850°C. For these samples, the PL emission peaks are centered at about 450 nm, 490 nm, and570 nm. The inset shows the PL lifetimes, which arise from a combination of radiative recombination (e.g., spontaneous emission) and non-radiative recombination. The PL decay was fitted using the following equation:

$$P(t) = P_1(t)e^{\left(-\frac{t}{\tau_1}\right)} + P_2(t)e^{\left(-\frac{t}{\tau_2}\right)} + P_0$$

For sample 1, $\tau_1$=2.4 ns, $\tau_2$=15.1 ns, and $P_2/P_1$=1.12; for sample 2, $\tau_1$=2.5 ns, $\tau_2$=15.8 ns, and $P_2/P_1$=0.85; and for sample 3, $\tau_1$=1.1 ns, $\tau_2$=8.1 ns, and $P_2/P_1$ =0.088. The PL decay curves reveal at least two distinct lifetimes, indicating the presence of multiple emission mechanisms. Notably, Sample 3 exhibits a faster recombination rate, primarily due to an increase in non-radiative recombination processes. This highlights the complex interplay between material properties and growth conditions in determining the optical behavior of SiN films, particularly their emission characteristics in the visible and SW-NIR spectrum.

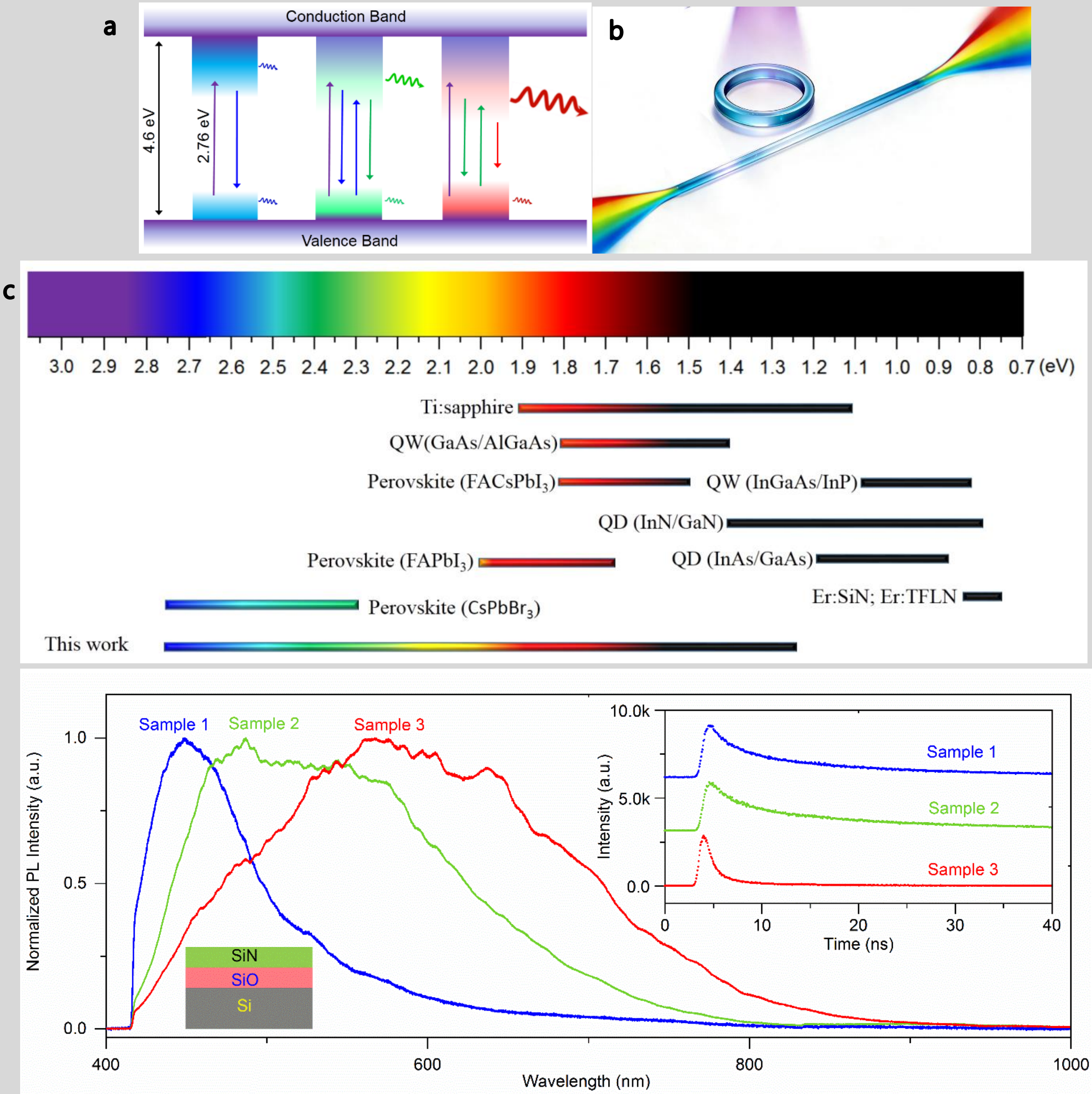

**Fig. 1 | Schema of e-h radiative and non- radiative recombination in SiN, light emission from SiN ring, along with a comparison of SiN ring radiative spectral bandwidths against those of other gain materials.** (a) Schematic illustration of radiative and non-radiative recombination in SiN with different ratio of Si to N. (b) Schematic illustration of broadband emission from a SiN ring. (c) Comparison of gain ranges between this work and other materials: Gain ranges

of GaAs/AlGaAs quantum wells (1.4–1.8 eV or 650–850 nm), InGaAs/InP quantum wells (0.8–1.1 eV or 1150–1600 nm) , InAs/GaAs quantum dots (0.9–1.2 eV or 1050–1350 nm), InN/GaN quantum dots (0.8–1.4 eV or 900–1550 nm), $FACsPbI_3$ perovskite (1.7–2.0 eV or 620–720 nm), $FaPbI_3$ perovskite (0.8–1.4 eV or 750–850 nm), 4.3 nm $CsPbBr_3$ perovskite quantum wires (2.3–2.8 eV or 450–540 nm) , erbium-doped SiN (Er:SiN) or Er:TFLNOI (0.75–0.85 eV or 1450–1650 nm), Ti:sapphire (1.1–1.9 eV or 650–1100 nm), and this work (1.24–2.75 eV or 450–1000 nm). (d) Normalized PL spectra of three SiN films deposited on $SiO_2$/Si substrates with varing $NH_3$ flow rates. Inset: Time resolved PL decay curves. Biexponential fiting reveals two distinct lifetime components for each sample. Sample 1 exhibits lifetimes $\tau_1$=2.4 ns (short) and $\tau_2$=15.1 ns (long). Sample 1 shows $\tau_1$=2.5 ns (short) and $\tau_2$=15.8 ns (long); Sample 3 demonstrate $\tau_1$=1.1 ns (short) and $\tau_2$=8.1 ns (long).

## SiN waveguide amplification

Fig. 2a illustrates fabrication process of SiN nanostructures. The fabrication process involves electron beam lithography (EBL) to pattern the nanostructures on the SiN film, followed by inductively coupled plasma (ICP) etching to transfer the pattern into the SiN layer.

Initial characterization of three SiN samples revealed distinct PL properties (Fig. 1). Sample 1 exhibited particularly strong green-band emission with minimal absorption at the 405 nm pump wavelength, making it optimal for waveguide amplifier fabrication. The selected waveguide featured a 2 μm × 800 nm cross-section with 2 mm length, embeded in $SiO_2$ and designed to support effective mode area 0.82 $\mu m^2$ at 405 nm and 0.86 $\mu m^2$ at 532.3 nm respectively.

For amplification testing, a nanosecond pulses signal centered at about 532.3 nm was co-propagated with the 405 nm pump through tapered fiber couplers (Extended Data Fig. 1). Fig. 2b demonstrates clear signal enhancement, achieving a net gain of 2.3× at 1.1 mW and 3.6× at 2.0 mW coupled pump power. This demonstrates the visible-light amplification in SiN waveguides.

## PL of SiN nanosturctures

The emission properties of SiN nanostructures were investigated using 300 nm-thick films patterned into one-dimensional photonic crystal (1D PhC) structures on silicon oxide substrates. While such substrate-bound PhCs typically exhibit weaker optical confinement than suspended structures due to asymmetric mode profiles, complete embedding in silicon oxide through PECVD of a 1 μm-thick $SiO_2$ cladding layer restored localized optical properties [17]. Optical characterization using 405 nm laser excitation revealed that a 500 nm-wide waveguide with elliptical nanoholes (300 nm major axis, 100 nm minor axis) and 210 nm periodicity produced a sharp emission peak at 625 nm with 0.8 nm linewidth (Q ≈ 780) (Fig. 2c). Increasing the period to 250 nm while maintaining identical geometry resulted in a redshifted emission at 749 nm with comparable linewidth (0.8 nm) and improved Q-factor (≈930), demonstrating >100 nm spectral tunability at visible spectrum through simple geometric modifications - a range challenging to achieve with conventional semiconductor gain materials.

Further development of distributed feedback (DFB) lasers using oxide-embedded SiN structures yielded narrow emission lines at 688 nm (Q ≈ 1,300) and 738 nm (Q ≈ 1,400) for 210 nm and 250 nm periods respectively, both with 0.5 nm linewidths (Fig. 2d). The observed fluorescence background, attributed to the large collection area of the measurement setup, does not compromise the demonstrated performance advantages of these integrated structures.

## PL of SiN ring resonators

In nanophotonic devices, ring resonators are widely utilized for their excellent lasing properties. Strong optical confinement in the waveguide enables ring resonators to achieve high Qs even at small radii.

Fig. 3a illustrates the $TE_0$ and $TE_1$ mode distributions for wavelengths of 450 nm and 1000 nm in a SiN waveguide with a width of 1 μm, or 2 μm, and a height of 300 nm. As shown in the figure, the 1 μm-wide waveguide provides strong confinement for the $TE_0$ mode at 450 nm, but the confinement significantly weakens for the $TE_0$ mode at 1000 nm due to the more than twofold increase in wavelength. To enhance confinement, the waveguide width was increased to 2 μm, which improved the confinement for both 450 nm and 1000 nm $TE_0$ modes. However, this also introduced higher-order modes, such as the $TE_1$ mode, leading to enhanced confinement for these modes as well.

A 12-μm-diameter ring resonator (1-μm width) was fabricated using the material from Sample 3 (Fig. 1), directly coupled to a 600-nm-wide waveguide (0 nm gap) without $SiO_2$ cladding. Experimentally, emission peaks initiates near 600 nm and terminates above 750 nm as cumulative losses exceed gain compensation. Although there is sufficient Qs below 600 nm and Fig.1 shows that the material's gain activation begins at about 500 nm, the strong absorption and non-radiative recombination in this spectral range precludes notable emission. Cold-cavity simulations reveal a strong wavelength-dependent Q factor: it decline sharply at longer wavelengths due to weakened mode confinement and proximity-induced coupling losses. Notably, despite sufficient gain above 750 nm, the sharply increased loss precludes notable emission.

Building on this platform, 16-μm-diameter ring resonator was fabricated using the gain medium from Sample 1 (Fig. 1), with a 300-nm-wide waveguide, 40 nm gap between the ring and output waveguide, exposed in air. This configuration produced sharp emission at 450 nm but exhibited complete spectral quenching beyond 650 nm—a behavior traced to the fundamental gain limitations of the material system. While Sample 1 demonstrates measurable gain at 600 nm (Fig. 3c), its rapid decline at longer wavelengths prevents emission despite maintained optical confinement. The achieved operational window (450–650 nm) directly correlates with the material's gain spectrum, revealing a critical size and gain dependent tradeoff in SiN ring resonators: increased cavity dimensions enhance mode confinement at longer wavelengths but cannot overcome gain limitations across broader spectral ranges. These findings establish material gain engineering as equally crucial as optical confinement design for realizing broadband integrated lasers in wide-band photonic platforms.

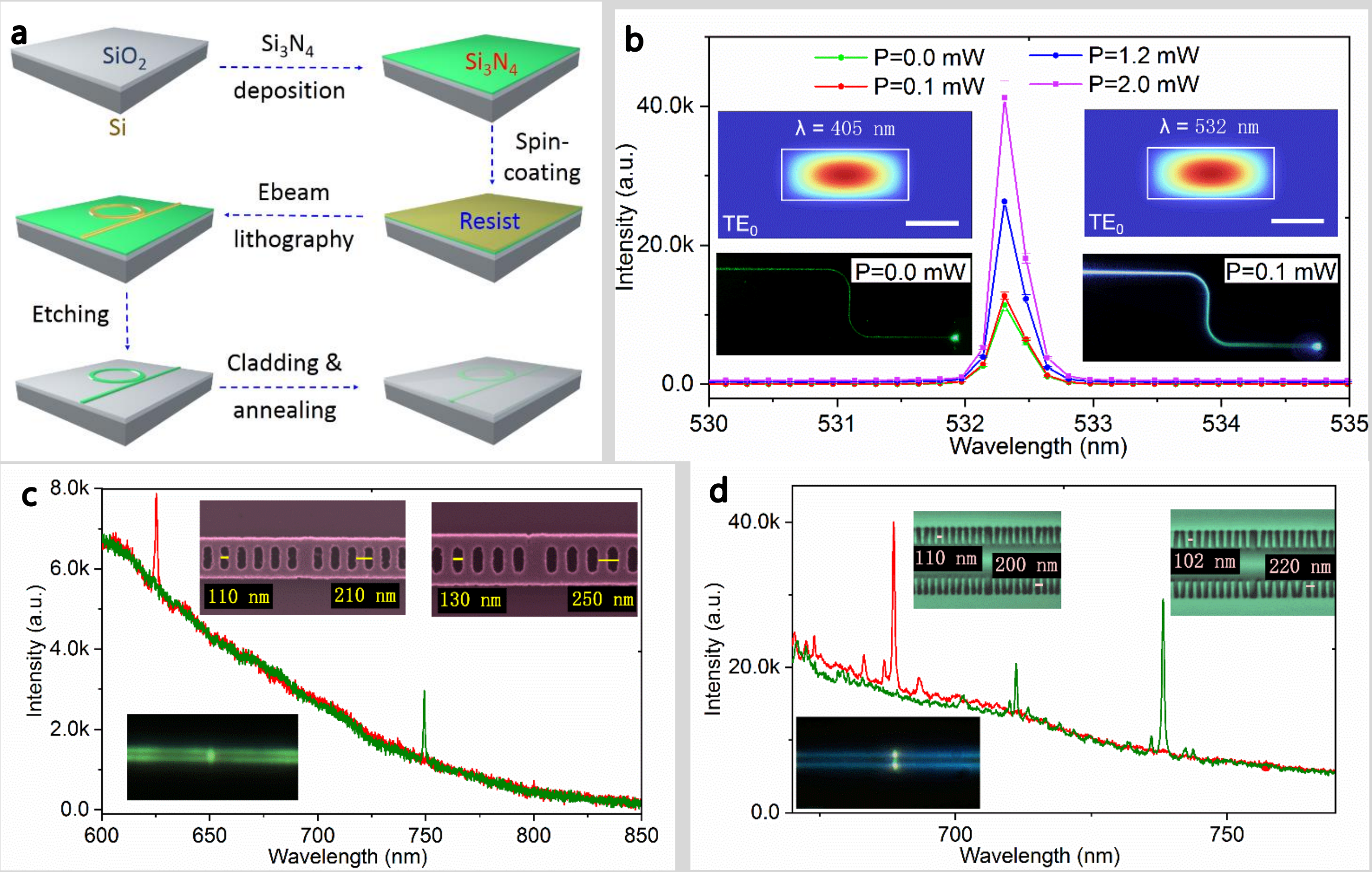

**Fig. 2 | Fabrication, PL, and amplification of SiN nanostructures and waveguides.** (a) Schematic of the fabrication process for SiN nanostructures. (b) Amplification in the SiN waveguide. The green and blue curves represent the emission spectra without and with a 1.5 mW blue light pump, respectively. The inset shows optical mode distributions in the waveguide with pump wavelength 405 nm and signal wavelength 532.3 nm, and optical images of the waveguide emission under without and with 405 nm pump conditions. Error bars represent standard deviation from four measurements. (c) PL spectra of SiN PhC structures with periods of 180 nm and 220 nm. The inset displays false-colored scanning electron microscopy (SEM) images and corresponding optical fluorescence images. (d) PL spectra of SiN DFB structures with periods of 180 nm and 220 nm, including a λ/4 phase shift. The inset shows false-colored SEM images and corresponding optical fluorescence images.

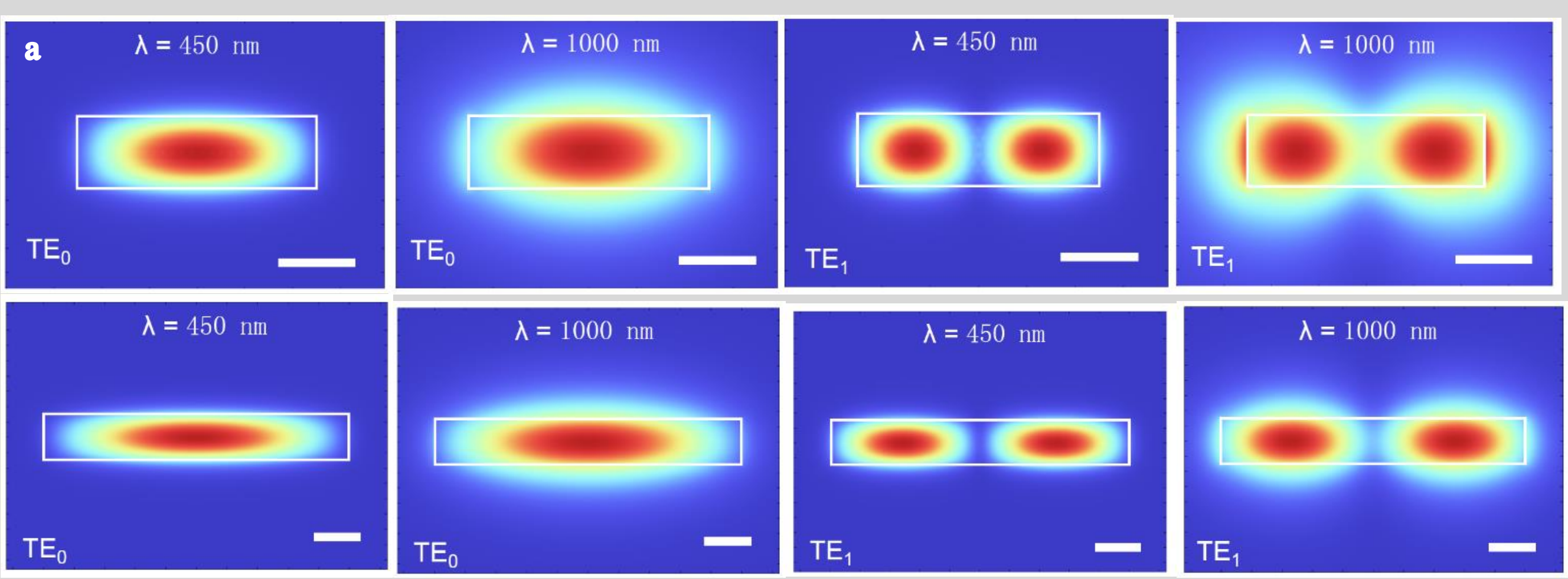

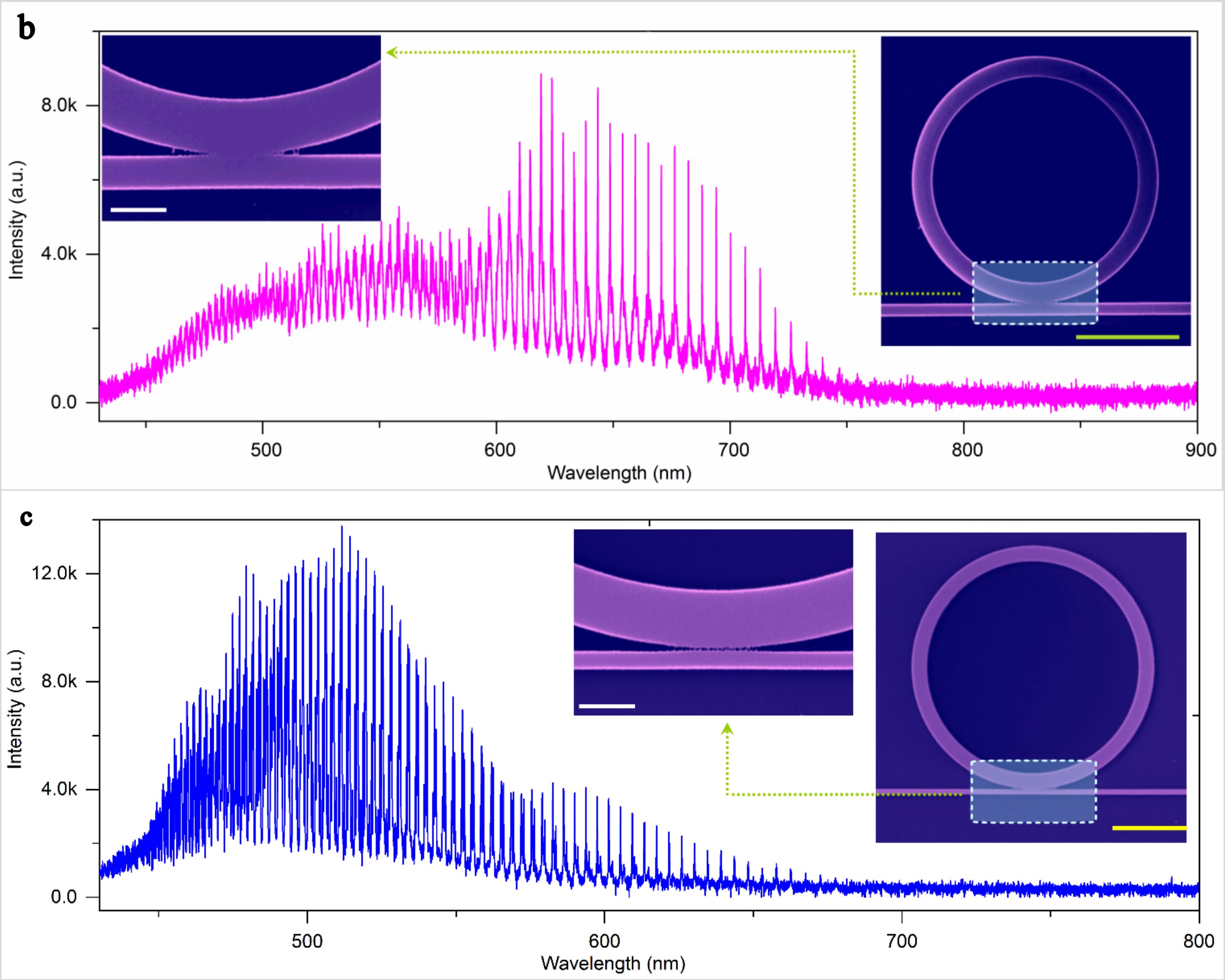

**Fig. 3 | Mode analysis and photon emission spectra from different types of chip-integrated SiN ring resonators.** (a) Mode analysis of SiN waveguides in SiO2 with a height of 300 nm and widths of 1 μm and 2 μm, respectively. Scale bar: 300 nm. (b, c) Emission intensity of the SiN ring resonator. Insets: False-colored SEM images of the ring resonator and magnified view of the coupling region between the micro-ring and bus waveguide, with 0 nm and 40 nm gap. Yellow and wihte scale bars in (b) and (c) are 5 μm and 1 μm respectively.

## Ultrawide-band SiN ring lasers

For ultrawide-band emission, spectral characterization required adaptive filtering to mitigate dispersion effects from broadband operation. A dual-filter system addressed transmission nonlinearities (e.g., 415 nm long-pass filter's residual blocking transmission characteristics at around 830 nm), with data fusion preserving spectral fidelity (Fig. 4a). Grating selection (1800 vs. 600 lines/mm) balanced resolution and range for full spectral capture—larger cavities suffered modal overlap at shorter wavelengths due to reduced free spectral range, necessitating enhanced resolution. While the 1800 lines/mm grating provided measurement up to 940 nm, the 600 lines/mm grating enabled extended wavelength range detection.

To broaden the emission spectral range, SiN ring resonators was implemented using Sample 2 — a material exhibiting PL comparable to Sample 1 but retaining marginal optical gain beyond 650 nm. A 12-μm-diameter ring resonator (1 μm waveguide width) coupled to a 400-nm-wide waveguide (100 nm gap) demonstrated emission extending to 900 nm (Fig. 4c), significantly surpassing the performance of previous configurations (Fig. 3b). This extended operation arises from resonant wavelength-selective reabsorption: shorter-wavelength light is partially absorbed to amplify longer wavelengths, compensating for the material's inherently weak SW-NIR gain (Fig. 1). The resultant spectrum deviates markedly from Sample 2's PL profile, highlighting cavity-enhanced spectral reshaping.

Scaling the resonator diameter to 16 μm (1 μm waveguide width, 50 nm gap) enabled emission extending to 940 nm (Fig. 4d) and distinct lasing threshold behavior under variable pumping (4.2–5.5 kW/cm², Fig. 4e). Selective mode excitation emerged at critical power densities, with a ×1.3 intensity scaling factor revealing threshold bifurcation between modes at different wavelength [18]. Further expansion to a 36-μm-diameter resonator (2 μm waveguide width, 50 nm gap) achieved record broadband emission (450 – 1000 nm, Fig. 4f, Extended Data Fig. 2c), exceeding conventional solid-state laser bandwidths. Theoretical analysis confirms this diameter-dependent performance enhancement: larger cavities maintain high SW-NIR Q-factors through improved mode confinement, while smaller counterparts (12–16 μm) exhibit rapid intensity decay beyond 900–1000 nm due to cumulative losses.

The system's tunability was further demonstrated through thermoelectric cooling (TEC) of a 12-μm resonator, achieving mode-hop-free tuning over 1.6 nm (centered at 738 nm) across a 70°C temperature range (Fig. 4b). This thermal tunability, combined with cavity-engineered spectral broadening, underscores the platform's potential for reconfigurable photonic systems.

Achieving broadband operation in SiN resonators necessitates meticulous balancing of gain-medium spectral response, cavity geometry-dependent loss mechanisms, higher-order mode interactions across wavelengths, and waveguide coupling efficiency. The interplay between these factors dictates device performance: the gain medium's wavelength-dependent emission must compensate for geometry-induced losses, while higher-order mode competition — particularly pronounced at shorter wavelengths in larger cavities — requires tailored suppression strategies to maintain spectral coherence. Simultaneously, optimized waveguide coupling ensures efficient light extraction without perturbing the delicate balance between gain and loss. Notably, while higher-order modes are traditionally suppressed in narrowband lasers, their controlled coexistence in broadband SiN resonators may enable novel applications in multimodal optical computing or processing, and high-density wavelength-division multiplexing (WDM), applications traditionally requiring complex hybrid systems.

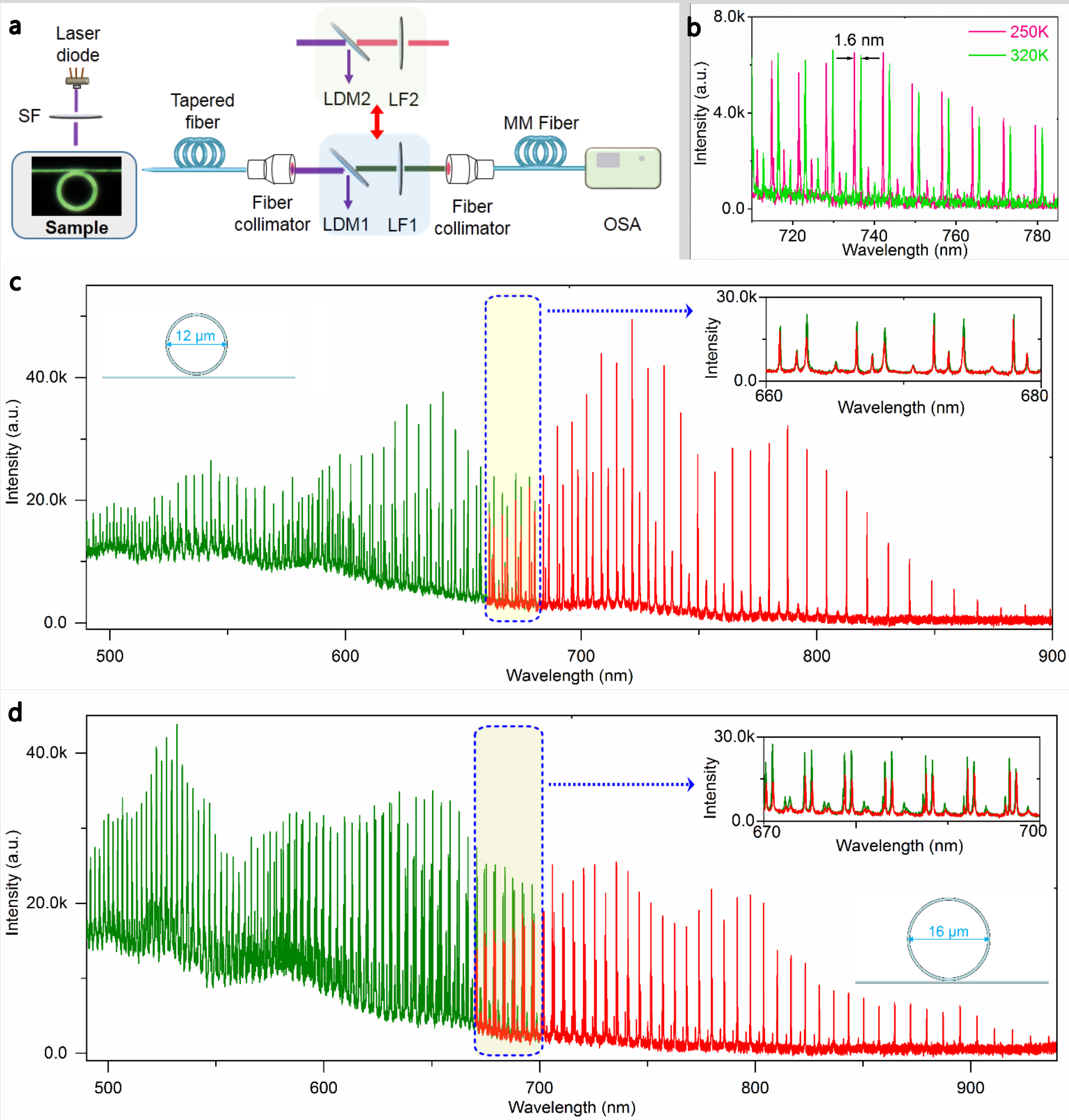
a
Laser diode
SF
Sample
Tapered fiber
Fiber collimator
LDM2
LF2
LDM1
LF1
MM Fiber
Fiber collimator
OSA
b
Intensity (a.u.)
8.0k
4.0k
0.0
1.6 nm
250K
320K
720
740
760
780
Wavelength (nm)
c
Intensity (a.u.)
40.0k
20.0k
0.0
12 μm
30.0k
Intensity
0.0
660
680
Wavelength (nm)
500
600
700
800
900
Wavelength (nm)
d
Intensity (a.u.)
40.0k
20.0k
0.0
30.0k
Intensity
0.0
670
700
Wavelength (nm)
16 μm
500
600
700
800
900
Wavelength (nm)

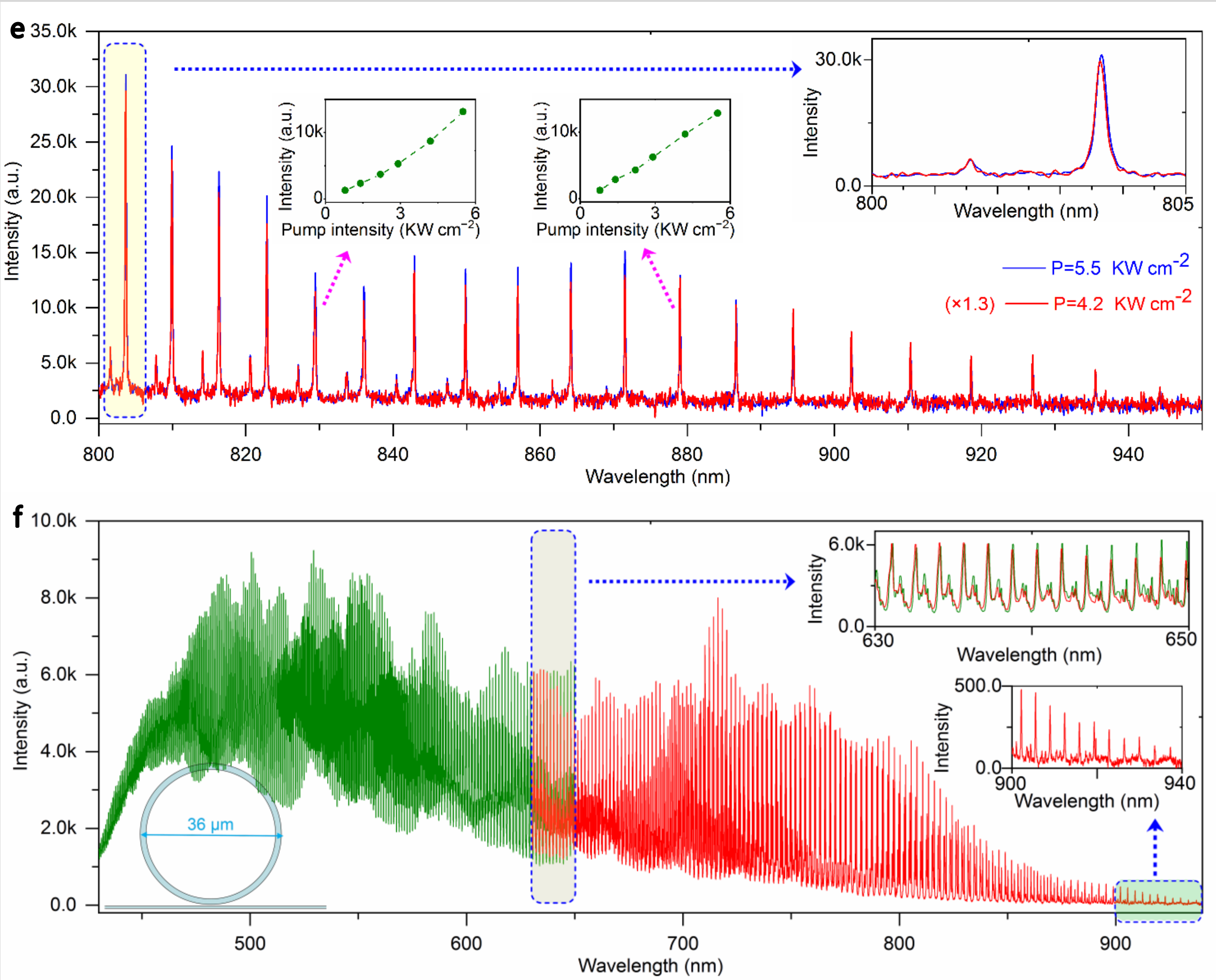

**Fig. 4 | Broadband integrated SiN ring lasers.** (a) Schematic of the experimental setup for measuring emission from SiN rings. Optical components include: SF, short-pass filter transmitting light below 415 nm; LDM1, long-pass dichroic mirror reflecting light below 415 nm and transmitting light above 415 nm; LF1, long-pass filter transmitting light above 415 nm; LDM2, long-pass dichroic mirror reflecting light below 540 nm and transmitting light above 540 nm; LF2, long-pass filter transmitting light above 540 nm. (b) Emission wavelength tuning of the 12-μm diameter ring resonator placed on a TEC. (c) Emission spectrum of a 12 μm-diameter SiN ring coupled to a 400 nm-wide waveguide with a 100 nm gap. Inset: Emission spectrum in the range of 660–680 nm. (d) Emission spectrum of a 16 μm-diameter SiN ring coupled to a 400 nm-wide waveguide with a 50 nm gap. Inset: Emission spectrum in the range of 670–700 nm. (e) Emission spectra of the 16 μm-diameter SiN ring at pump powers of 5.5 kW/cm² and 4.2 kW/cm² (scaled by a factor of 1.3). Inset: Output emission peak power versus pump power at 829 nm and 879 nm, and magnified view from 800nm to 805 nm. (f) Emission spectrum of a 36 μm-diameter SiN ring resonator coupled to a 300 nm-wide waveguide with a 50 nm gap, with backgound noise smoothed. Inset: Detailed emission spectra in the ranges of 630–650 nm and 900–940 nm. All the SiN ring resonators materials are the same with that of Sample 2 shown in Fig.1, and buried in $SiO_2$. The green and red lines in (c), (d) and (f) are PL from the different filters system respectively as shown in (a).

## Discussion

This work establishes SiN as a monolithically integrable broadband gain medium spanning 450–1,000 nm, overcoming the spectral limitations of Ti:sapphire systems while addressing critical wavelength compatibility challenges in silicon photonics. The demonstrated visible-to-SW-NIR emission directly aligns with silicon's native photodetector responsivity (190 – 1100 nm) — as exploited in CMOS-integrated avalanche photodiodes [19,20] — enabling fully monolithic intra-chip optical interconnects with superior bandwidth density compared to conventional telecom-wavelength solutions constrained by III-V detector incompatibility. Beyond intra-chip communications, the extended visible and SW-NIR operation, provides capabilities for quantum optics, atomic and molecular physics, while the strong optical confinement (n = 2.0–2.2) enables sub-10-μm-radius high-Q resonators unattainable in low-index platforms like Ti:sapphire.

Device optimization reveals fundamental tradeoffs: broadband operation demands meticulous balancing of gain-bandwidth product management, short-wavelength reabsorption mitigation, and higher-order mode suppression—particularly critical in large-radius resonators where modal competition intensifies at shorter wavelengths. Paradoxically, these "undesirable" higher-order modes may enable WDM or multimodal optical computing architectures when strategically controlled [21, 22], expanding SiN's utility beyond conventional single-mode paradigms.

Structural engineering further enhances functionality, as evidenced by 1D photonic crystals achieving >100 nm tuning at visible spectrum through periodicity control, surpassing conventional semiconductors limited by fixed bandgaps. SiN's operational range was extended into the 520 nm regime previously exclusive to Ti:sapphire, with nitrogen stoichiometry adjustments suggesting future sub-500 nm operation.

The material's CMOS compatibility emerges as a pivotal advantage: LPCVD growth permits large area uniform 200–300 nm thick film without cracking adhering to foundry design rules. This positions SiN as a gateway to heterogeneous systems — monolithic "light-source + waveguide + detector" architectures leveraging silicon photodiodes. This approach not only bridges Ti:sapphire's spectral gaps but also unlocks applications spanning intra-chip communications, neural networks and atomic sensing, while maintaining full compatibility with other materials, like emerging 2D materials, lithium niobate photonics and $Al_2O_3$ photonics. It should be emphasized that by exploiting intrinsic material property variations within a single SiN platform, it can achieve heterogeneous integration-like functionality without complex hybrid schemes — a paradigm shift for integrated photonics. This work redefines SiN from a passive waveguide material to an active multifunctional platform, where engineered growth and cavity synergies enable unprecedented control over light-matter interactions in monolithically integrated systems.

## METHODS

### Device fabrication

The SiN nanodevices were fabricated on 2-μm-thick thermally grown silicon dioxide ($SiO_2$) substrates. 300-nm-thick and 800-nm-thick SiN layer was deposited via LPCVD using $SiH_2Cl_2$ and $NH_3$ precursor gases at flow ratios of 4:1 to 1:4 and temperatures of 750–850° C to optimize stoichiometry and optical properties. Nanostructures were patterned using high-resolution electron-beam lithography (EBL) followed by inductively coupled plasma reactive ion etching (ICP-RIE) with a $CHF_3$/$SF_6$ gas mixture, achieving anisotropic profiles with sidewall roughness <1 nm. Residual resist was removed via oxygen plasma ashing, and selected devices were encapsulated with a 1-μm-thick $SiO_2$ cladding layer deposited by plasma-enhanced chemical vapor deposition (PECVD) to enhance optical confinement.

### Device Design

Waveguide and resonator geometries were optimized through finite-difference time-domain (FDTD) simulations. Amplification waveguides were designed with 800 nm thickness to maximize modal overlap, while resonator structures—including 1D photonic crystals, Bragg mirrors, and ring resonators—employed 300 nm thickness for balanced optical confinement. Waveguide widths of 300 nm and 600 nm were implemented to support effective coupling across 450–1,000 nm, with ring resonator widths (1–2 μm) and coupling gaps (0–100 nm) tailored to specific spectral regimes.

### PL and lifetime mearment of SiN film

PL characterization utilized a 405 nm diode laser with a 415 nm bandpass filter for excitation, coupled to a time-correlated single-photon counting (TCSPC) system featuring a silicon single-photon avalanche diode (SPAD) detector (64 ps temporal resolution).

### Optical characterization

Optical amplification experiments employed a 405 nm single-mode pump laser diode and nanosecond signal laser centered at about 532.3 nm, combined through a semi-reflective mirror and coupled into SiN waveguides via fiber collimator lensed fibers (2.5 μm spot size, about 16% coupling efficiency). The nanosecond laser operates at a repetition rate of 4 KHz, with a pulse width of 3.9 ns and a single-pulse energy of 37 μJ.

For ring resonator characterization, a 405 nm multimode laser focused through a 100× objective (~100 μm spot) provided optical pumping, with waveguide-coupled emission collected via lensed fiber and filtered through a dual-filter system.

Emission spectra were acquired using a spectrometer with a thermoelectrically cooled CCD (−60 ° C) and three interchangeable gratings: 300 lines/mm (0.3 nm resolution, 350–1,100 nm broadband), 600 lines/mm (0.15 nm resolution, 350–1,100 nm), and 1800 lines/mm (0.05 nm resolution, 430–940 nm). The system's silicon CCD detector exhibited peak quantum efficiency between 400–1,000 nm with sensitivity sharply declining at longer wavelengths due to the silicon bandgap limit, enabling precise detection of weak

PL features and nanoscale spectral shifts critical for device optimization. All measurements incorporated dark current subtraction and flat-field calibration to ensure spectral fidelity.

## Acknowledgments

The author would like to acknowledge funding support from the Hunan Provincial Major Sci-tech Program (grant no. 2023ZJ1010).

## Competing interests

The authors declare no competing interests.

## Data Availability

All data are available from the corresponding author upon reasonable request.